\documentclass[12pt]{article}
\usepackage{amsmath,amssymb,amsfonts}
\usepackage{txfonts}
\usepackage[T1]{fontenc}
\usepackage{bm,booktabs,array,graphicx}
\usepackage[font=small,labelfont=bf]{caption}
\usepackage[numbers,sort&compress]{natbib}
\usepackage[colorlinks=true,citecolor=blue,linkcolor=blue,urlcolor=blue]{hyperref}
\usepackage{microtype}

\newcommand{\Dstar}{\Delta^{\!*}}
\newcommand{\psihat}{\hat{\psi}}
\newcommand{\R}{\mathbb{R}}
\newcommand{\norm}[1]{\left\|#1\right\|}

\title{Shape effects on the Shafranov shift of Solov'ev equilibria with prescribed boundaries}
\author{D.~Abate\textsuperscript{1}\thanks{Corresponding author: domenico.abate@igi.cnr.it}}

\date{November 2026}

\begin{document}
	\maketitle
	\begin{center}
		\textsuperscript{1}	Consorzio RFX (CNR, ENEA, INFN, Università di Padova,\\ Acciaierie Venete SpA), Corso Stati Uniti 4, 35127 Padova, Italy \\
		
	\end{center}
	
	\begin{abstract}
		The Shafranov shift, the outward displacement of the magnetic axis of a
toroidal plasma, is known analytically for circular and elliptical
cross-sections, whereas its dependence on shape in strongly shaped plasmas has
not been studied systematically. Here the shift is analyzed for a wide range
of prescribed boundaries, at fixed profiles or poloidal beta, using exact
Solov'ev solutions of the Grad--Shafranov equation with the boundary imposed in
a least-squares sense. For shaped boundaries the shift depends on the
point from which it is measured, usually the geometric center, the
midpoint of the radial extent of the boundary. Therefore, a straight-cylinder equilibrium with the same plasma shape is introduced as the zero-shift reference state, and the shift is measured from its magnetic axis. With this reference, some known features of the shift measured from the geometric center turn out to be effects of the reference point, whose error depends only on the boundary and
can also make the shift negative. Three boundary parametrizations are used to analyze positive and negative triangularity, squareness from square to concave shapes, and
non-convex star polygons of arbitrary symmetry order. At fixed poloidal beta
the shift depends only weakly on triangularity: the larger shift of negative than of positive triangularity plasmas obtained with the geometric center is
an effect of the reference point and nearly disappears with the cylindrical
reference. Negative squareness and concavity, instead, reduce the shift itself: for concave plasma cross-sections it is almost half that of elliptical and D-shaped ones.
	\end{abstract}
	
	%% =============================================================================
	\section{Introduction}
	\label{sec:intro}
	%% =============================================================================
	
	In toroidal magnetic confinement fusion devices, the magnetic axis of the plasma equilibrium does not coincide with the geometric axis, but is displaced outward. This displacement, the Shafranov shift $\Delta$, was calculated analytically in the large-aspect-ratio limit, retaining the toroidicity of the system in the linear approximation and prescribing the shape of the flux surfaces, as shifted circles~\cite{shafranov1963,shafranov1966} or shifted ellipses~\cite{pustovitov2000,pustovitov2021}. %In this limit the shift is of order $(a^2/2R_0)(\beta_p+l_i/2)$, where $a$ and $R_0$ are the minor and major radii, $\beta_p$ the poloidal beta and $l_i$ the internal inductance~\cite{wesson2011,freidberg2014}.
	The shift is a basic parameter of a toroidal equilibrium. It is commonly used as a metric for the position of the plasma core, it determines the equilibrium limit on the plasma pressure~\cite{freidberg2014,cerfon2010,pustovitov2000}, and it affects the magnetohydrodynamic stability~\cite{marinoni2021,song2021} and the confinement~\cite{marinoni2021,garbet2003} of tokamak plasmas; in stellarators it also affects neoclassical transport and fast-particle losses~\cite{helander2026}, and a small shift is a design goal of power-plant configurations~\cite{lion2025}.
	
	Since the proposal of tokamaks with non-circular cross-sections~\cite{artsimovich1972}, plasma shaping has been one of the main ways of extending the stability limits and improving the confinement of toroidal plasmas. Elongated and D-shaped cross-sections are now standard~\cite{wesson2011,freidberg2014}. Negative triangularity (NT) reduces turbulent transport and gives high confinement without edge-localised modes (ELMs)~\cite{camenen2007,marinoni2021,thome2024}, and it has been proposed for a fusion pilot plant~\cite{manta2024}. Squareness has been used to optimise stability and confinement~\cite{holcomb2009}, and high squareness is predicted to give access to ELM-free operation in spherical tokamaks~\cite{parisi2025}. Non-convex, bean-shaped cross-sections have been studied for high-beta operation~\cite{chance1983,bell1990}, and they are also found in stellarators~\cite{lion2025}.
	
	The classical theory, which prescribes the flux surfaces with one or two shape parameters~\cite{pustovitov2000}, does not apply directly to these strongly shaped plasmas, and strong shaping can change the shift substantially~\cite{helander2026}; for tokamaks, its dependence on shape is known mainly from numerical equilibria, such as those giving a larger shift for NT plasmas~\cite{marinoni2021,song2021}. In general, for shaped cross-sections the shift is not uniquely defined, because the center from which it is measured is not~\cite{rodriguez2023}. Thus, in tokamak practice it is measured from the geometric center of the boundary, the midpoint of its radial extent~\cite{sauter2016,marinoni2021}, which coincides with the center of the flux surfaces of the classical theory for cross-sections symmetric in the radial direction, such as circles and ellipses, but not for triangular or other asymmetric ones. For stellarators, an average measure of the shift relative to the zero-$\beta$ equilibrium with the same boundary has recently been introduced~\cite{helander2026}. Thus, the dependence of the shift on shape, measured from a reference defined consistently for all possible shapes, requires a systematic study.
	
	In this work, the Shafranov shift is studied as a function of the shape of the boundary, from D-shapes through NT to non-convex cross-sections. Analytical Solov'ev equilibria~\cite{solovev1968,cerfon2010} are used, with the boundary prescribed by a least-squares fit~\cite{aydemir2019} on a basis of homogeneous solutions generated to any degree by a recursion~\cite{zheng1996,aydemir2019}: the Grad--Shafranov equation is satisfied exactly and, at fixed profiles or at fixed $\beta_p$, shape is the only variable. Three parametrizations of the boundary are used as a progression in shape. Miller boundaries~\cite{miller1998} (Family~A) describe the passage from positive to negative triangularity. Luce boundaries~\cite{luce2013} (Family~B), built from a superellipse in each quadrant, add squareness as an independent parameter; below a critical squareness their sides become concave, so that the scan leads continuously from tokamak shapes to non-convex ones. Star polygons generated by the superformula of Gielis~\cite{gielis2003} (Family~C) provide the extreme case: they have no direct tokamak counterpart, but they extend the scan to strongly concave boundaries with a controlled symmetry order. The three-fold polygons are triangles with strong positive triangularity (up to $\delta=1$), and for odd symmetry order and strong concavity the inboard side is indented, as in the bean-shaped plasmas of PBX-M~\cite{bell1990}, so that the boundary is not symmetric in the radial direction, like the bean-shaped cross-sections of stellarators~\cite{lion2025}. Following large-aspect-ratio expansion~\cite{shafranov1963,wesson2011,freidberg2014}, the straight-cylinder equilibrium with the same cross-section is taken as the zero-shift reference state and the shift is measured from its magnetic axis; the errors introduced by the conventional center and by other natural choices, the barycenter of the boundary and the area and current centroids, are then quantified. Measured from the cylindrical reference, the shift is outward for every shape considered, whereas measured from the geometric center it can also become negative (for odd-fold star polygons and concave boundaries with positive triangularity). At fixed $\beta_p$ the shift depends only weakly on triangularity, so that the larger shift of NT plasmas found with the geometric center is an effect of the reference point, whereas negative squareness and concavity reduce the shift itself.
	
	The paper is organised as follows. Section~\ref{sec:solution} presents the model, the least-squares construction, the figures of merit and the reference state for the shift; Sections~\ref{sec:convex} and~\ref{sec:nonconvex} present the shaped tokamak boundaries and the star polygons, and Section~\ref{sec:conclusions} summarises the conclusions.
	
	%% =============================================================================
	\section{Mathematical model}
	\label{sec:solution}
	
	\subsection{Solov'ev profiles and analytical solution}
	A cylindrical coordinate system $(R,\varphi,Z)$ is adopted, with $R$ the major radius, $\varphi$ the toroidal angle and $Z$ the vertical coordinate. The magnetic field of an axisymmetric equilibrium is written as $\mathbf{B}=(F/R)\,\mathbf{e}_\varphi+(1/R)\,\nabla\Psi\times\mathbf{e}_\varphi$, where $2\pi\Psi$ is the poloidal flux and $F=RB_\varphi$~\cite{cerfon2010}. With normalized coordinates $x=R/R_0$, $y=Z/R_0$, $\psihat=\Psi/\Psi_0$,
	the Solov'ev choice of linear source functions~\cite{solovev1968},
	\begin{equation}
		-\frac{\mu_0 R_0^4}{\Psi_0^2}\frac{dp}{d\psihat} = 1-A,
		\qquad
		-\frac{R_0^2}{\Psi_0^2}F\frac{dF}{d\psihat} = A,
		\label{eq:solovev}
	\end{equation}
	converts the Grad--Shafranov (GS) equation~\cite{grad1958,shafranov1957} into the linear inhomogeneous problem
	\begin{equation}
		\Dstar\psihat \equiv
		\frac{\partial^2\psihat}{\partial x^2}
		-\frac{1}{x}\frac{\partial\psihat}{\partial x}
		+\frac{\partial^2\psihat}{\partial y^2}
		= A + (1-A)\,x^2.
		\label{eq:GSnorm}
	\end{equation}
	The parameter $A$ sets the relative weight of the two source terms in the toroidal current density,
	\begin{equation}
		j_\varphi
		= -\frac{\Psi_0}{\mu_0 R_0^3}
		\Bigl[\frac{A}{x}+ (1-A)x\Bigr].
		\label{eq:jtor}
	\end{equation}
	Inside the plasma $j_\varphi$ depends only on $x$ while the shape of boundary enters through the domain on which $j_\varphi$ is non-zero. The term $A/x$, arising from $FF'$, peaks on the high-field side, and the term $(1-A)x$, arising from $dp/d\psihat$, peaks on the low-field side.
	A pressure decreasing from the axis to the boundary requires $A<1$; $A=1$ gives a force-free equilibrium ($p'=0$, $\beta=0$) and $A=0$ gives $FF'=0$, a vacuum toroidal field~\cite{cerfon2010}. Integrating Eq.~\eqref{eq:solovev} gives $F^2=F_b^2-2A\Psi_0^2\psihat/R_0^2$~\cite{cerfon2010}, with $F_b=R_0B_0$ the value on the boundary, where $\psihat=0$; since $\psihat<0$ inside the plasma, $A>0$ ($A<0$) makes the toroidal field larger (smaller) than the vacuum field, that is, the plasma is paramagnetic (diamagnetic). In this work $A\in(0,1)$, so that the two terms in Eq.~\eqref{eq:jtor} have the same sign and the current density does not reverse.
	The cases considered in Secs.~\ref{sec:convex} and~\ref{sec:nonconvex} are computed with $A=0.11$, which gives
	$\beta_p\approx0.86$--$0.91$ for the convex tokamak boundaries and $\beta_t\approx2$--$3\,\%$, representative
	of conventional aspect-ratio tokamak operation at modest beta. For a given boundary $A$ is the only free parameter of the profiles: it enters the particular solution $\psi_p$ of Eq.~\eqref{eq:decomp}, and hence the boundary system of Sec.~\ref{sec:method}, and it sets $\beta_p$ through Eq.~\eqref{eq:betap}. In the scans at fixed $\beta_p$ of Secs.~\ref{sec:convex} and~\ref{sec:nonconvex}, instead, it is adjusted for each shape to obtain the prescribed value. In general, the choice of $A$ can be made to target different physical regimes~\cite{cerfon2010}.
	
	The general solution splits into a particular solution and a combination of $N_b$
	homogeneous ones,
	\begin{equation}
		\psihat(x,y) = \psi_p(x,y;A) + \sum_{i=1}^{N_b}c_i\,\psi_i(x,y),
		\qquad
		\psi_p = \frac{A}{2}\,y^2 + \frac{1-A}{8}\,x^4,
		\label{eq:decomp}
	\end{equation}
	where $\Dstar\psi_i=0$ and $\psi_p$ satisfies Eq.~(\ref{eq:GSnorm}) by direct computation. The toroidal current density follows from the flux as $\mu_0Rj_\varphi=-\Dstar\Psi$. Since $\Dstar\psi_i=0$, the homogeneous solutions carry no current: the current density of Eq.~\eqref{eq:jtor} comes entirely from $\psi_p$, whereas the homogeneous solutions add current-free (vacuum) poloidal fields, whose coefficients $c_i$ are chosen so that the prescribed boundary is a flux surface (Sec.~\ref{sec:method}).
	
	The homogeneous solutions are generated as in~\cite{zheng1996,aydemir2019}. All the boundaries considered in this work are up-down symmetric, and each homogeneous solution is expanded in even powers of $y$,
	\begin{equation}
		\psi_i(x,y)=\sum_{j=0}^{\infty}f_j(x)\,y^{2j},
		\label{eq:expansion}
	\end{equation}
	where the first coefficient, $f_0(x)=\psi_i(x,0)$, is the profile of the solution on the midplane $y=0$. Substituting Eq.~\eqref{eq:expansion} in $\Dstar\psi_i=0$ and collecting equal powers of $y$ gives the recursion (Eq.~(7) of~\cite{aydemir2019})
	\begin{equation}
		f_{j+1}=-\frac{f_j''-f_j'/x}{(2j+1)(2j+2)},
		\label{eq:recursion}
	\end{equation}
	which determines all the coefficients, and hence the whole solution, from its midplane profile $f_0$. Each step requires only the derivatives: for a power of $x$ and for a power multiplied by $\ln x$,
	\begin{equation}
		\begin{aligned}
			f&=x^n: &\qquad f''-\frac{f'}{x}&=n(n-2)\,x^{n-2},\\
			f&=x^n\ln x: &\qquad f''-\frac{f'}{x}&=n(n-2)\,x^{n-2}\ln x+2(n-1)\,x^{n-2},
		\end{aligned}
		\label{eq:Lpow}
	\end{equation}
	since both $f''$ and $f'/x$ lower the power of $x$ by two. At each step of Eq.~\eqref{eq:recursion} the power of $x$ therefore decreases by two, while the power of $y$ in Eq.~\eqref{eq:expansion} increases by two: for $f_0=x^{2k}$ the terms of the series are proportional to $x^{2k}$, $x^{2k-2}y^2$, $x^{2k-4}y^4,\dots$, all of degree $2k$ in $x$ and $y$. Starting from $f_0=x^{2k}$ with $k\geq1$, after $k-1$ steps the coefficient is proportional to $x^2$, for which Eq.~\eqref{eq:Lpow} gives zero, and the series ends; starting from $f_0=x^{2k}\ln x$, it ends one step later, after $x^2\ln x$ has produced a constant. The infinite series~\eqref{eq:expansion} thus reduces to a finite sum, a polynomial or a polynomial with logarithmic terms of degree $2k$, which is an exact solution of $\Dstar\psi=0$. For instance, $f_0=x^4$ gives $f_1=-4x^2$ and $f_2=0$, that is, $\psi=x^4-4x^2y^2$. The seed $f_0=1$ ($k=0$) gives the constant solution, whereas the logarithmic seed $\ln x$ would generate negative powers of $x$ and is not used.
	
	The basis is formed by these solutions up to a maximum degree $2K$: the constant and, for each $k=1,\dots,K$, the polynomial and the logarithmic solution of degree $2k$, for a total of $N_b=2K+1$ functions. For $K=3$ it is the same basis of~\cite{cerfon2010}, with $N_b=7$ functions for up-down symmetric equilibria (twelve for up-down asymmetric ones). The maximum degree sets the angular resolution of the basis: in polar coordinates $(r,\vartheta)$ centered at $(x,y)=(1,0)$, the polynomial solution of degree $2k$ contains poloidal harmonics up to $\cos 2k\vartheta$. A few functions therefore suffice for smooth elongated and triangular boundaries, whereas vertices of infinite curvature, concave sides and high orders of symmetry, whose description needs high harmonics, require a higher degree. The values of $N_b$ used for each family are given in Secs.~\ref{sec:convex} and~\ref{sec:nonconvex}.
	A crucial property of Eq.~(\ref{eq:decomp}) is that the GS equation is
	satisfied \emph{identically} for any choice of $c_i$: the coefficients
	serve only to position the zero-flux surface $\psihat=0$ at the desired
	plasma boundary.
	
	%% =============================================================================
	\subsection{Least-squares determination of LCFS coefficients}
	\label{sec:method}
	%% =============================================================================
	
	In the classical construction~\cite{cerfon2010} the $N_b$ coefficients are fixed by conditions at a few boundary points, so that the shape of the last closed flux surface (LCFS) is an output. Following~\cite{aydemir2019,li2008}, the condition $\psihat=0$ is instead enforced at $M>N_b$ points of the prescribed boundary and the resulting overdetermined system is solved in a least-squares sense.
	Given a plasma boundary sampled at $M$ points $\theta_j=2\pi j/M$,
	the condition $\psihat=0$ on the LCFS gives the linear system
	$\mathbf{G}\mathbf{c}=\mathbf{f}$,
	\begin{equation}
		G_{ji} = \psi_i(x_j,y_j), \qquad
		f_j    = -\psi_p(x_j,y_j;A),
		\label{eq:linsys}
	\end{equation}
	with $\mathbf{G}\in\R^{M\times N_b}$.
	For $M\gg N_b$ the system is overdetermined and has in general no exact solution. The problem is linear in $\mathbf{c}$: the coefficients are taken as the linear least-squares solution, which minimises the norm of the residual $\mathbf{G}\mathbf{c}-\mathbf{f}$,
	\begin{equation}
		\mathbf{c}^* = \operatorname{argmin}_{\mathbf{c}}
		\norm{\mathbf{G}\mathbf{c}-\mathbf{f}}_2
		\label{eq:lstsq}
	\end{equation}
	and is given by $\mathbf{c}^*=\mathbf{G}^{+}\mathbf{f}$, with $\mathbf{G}^{+}$ the pseudo-inverse of $\mathbf{G}$ computed by singular-value decomposition~\cite{golub2013} ($M=600$ in all cases).
	The quality of the boundary fit is quantified by the relative residual
	\begin{equation}
		\eta = \frac{\norm{\mathbf{G}\mathbf{c}^*-\mathbf{f}}_2}
		{\norm{\mathbf{f}}_2}.
		\label{eq:eta}
	\end{equation}
	Here and in the following $\eta$, as well as $\eta_\psi$ of Eq.~\eqref{eq:etapsi}, is expressed in percent. Cases with $\eta<0.5\,\%$ are accepted: in the scans they have smooth flux surfaces, whereas for $\eta\gtrsim0.7\,\%$ the LCFS is visibly distorted. The Miller boundaries of Family~A give $\eta\leq10^{-3}\,\%$, and all Family~A and~B boundaries $\eta<0.15\,\%$.
	
	The residual $\eta$ is a root-mean-square measure, so that an error concentrated at a few points, such as the vertices of a polygon, can leave it small. The local error is therefore measured by the largest value of $|\psihat|$ on the prescribed boundary, in units of the flux difference between the magnetic axis and the LCFS,
	\begin{equation}
		\eta_\psi=\frac{\max_{\partial\Omega}|\psihat|}{|\psihat_\text{ax}|}.
		\label{eq:etapsi}
	\end{equation}
	In terms of the normalized flux $\psi_N=1-\psihat/\psihat_\text{ax}$, equal to 0 at the axis and 1 on the LCFS, the prescribed boundary lies, at its worst point, on the surface $\psi_N=1\pm\eta_\psi$, inside the plasma for the minus sign. For instance, the star polygon with $m=4$, $n=0.65$ has $\eta=0.30\,\%$ but $\eta_\psi=44.5\,\%$.
	For Family~A, $\eta_\psi<0.01\,\%$; for Family~B, $\eta_\psi\leq7.8\,\%$; for Family~C, $\eta_\psi\leq8.2\,\%$ in 38 of the 45 valid cases. In the remaining seven, strongly concave or many-cornered cases ($11\,\%\leq\eta_\psi\leq45\,\%$), the zero-flux contour cuts off the vertices of the prescribed boundary, so that the computed LCFS differs from the prescribed one near the vertices. These cases are kept, but they are marked with a dagger in Tables~\ref{tab:fom} and~\ref{tab:err}, and their results should be regarded as approximate. As a check of the method, the NSTX-like equilibrium of~\cite{cerfon2010} is reproduced with $\beta_p=1.066$ and $\beta_t=16.2\,\%$, against $1.07$ and $16\,\%$.
	
	\subsection{Figures of merit}
	The basic properties of Solov’ev MHD equilibria are usually described by a few figures of merit (FOM) \cite{cerfon2010}. Substituting the Solov'ev profiles into the definitions of
	$\beta_p=2\mu_0\langle p\rangle/\bar{B}_p^2$ and
	$\beta_t=2\mu_0\langle p\rangle/B_0^2$, the normalization constant
	$\Psi_0$ cancels and one obtains~\cite{cerfon2010}
	\begin{equation}
		\beta_p = \frac{2(1-A)C_p^2\hat{P}}{\hat{J}^2\hat{V}},
		\qquad
		\beta_t = \frac{\beta_p\,\varepsilon^2}{q_*^2},
		\label{eq:betap}
	\end{equation}
	where $C_p=\oint dl/R_0$ and
	\begin{equation}
		\hat{V} = \iint x\,dx\,dy, \quad
		\hat{J} = \iint \frac{A+(1-A)x^2}{x}\,dx\,dy, \quad
		\hat{P} = -\iint \psihat\,x\,dx\,dy > 0.
		\label{eq:integrals}
	\end{equation}
	For shaped tokamak plasmas the Shafranov shift is commonly measured from the geometric center of the plasma boundary~\cite{marinoni2021,song2021},
	\begin{equation}
		\frac{\Delta_\text{geom}}{a} = \frac{x_\text{ax} - x_\text{geom}}{\varepsilon},
		\label{eq:shafranov}
	\end{equation}
	where $x_\text{ax}$ is the radial position of the magnetic axis,
	$x_\text{geom}=(x_\text{max}+x_\text{min})/2$ is the geometric
	center of the LCFS~\cite{sauter2016} and $\varepsilon=a/R_0=(x_\text{max}-x_\text{min})/2$ is the inverse aspect ratio~\cite{luce2013}; $\Delta_\text{geom}$ is referred to as the conventional shift in the following.
	For cross-sections symmetric in the radial direction, such as the circles and ellipses of the shifted-surface models~\cite{shafranov1966,pustovitov2000,pustovitov2021}, Eq.~\eqref{eq:shafranov} coincides with the classical definition of the Shafranov shift, the displacement of the centers of the flux surfaces; for triangular and other asymmetric cross-sections the center is not uniquely defined~\cite{rodriguez2023}, and the choice of the reference is discussed next.
	
	\subsection{Reference state for the Shafranov shift}
	\label{sec:center}
	At large aspect ratio the Grad--Shafranov equation reduces, at leading order in $\varepsilon$, to the equilibrium of a straight cylinder with the same cross-section, and the Shafranov shift is the displacement of the magnetic axis produced by the first-order, toroidal terms~\cite{shafranov1963,wesson2011,freidberg2014}. In analogy with the zero-$\beta$ reference state used for stellarators~\cite{helander2026}, the cylindrical equilibrium is therefore taken here as the zero-shift reference state. For the Solov'ev source the current density of the cylinder is uniform for any $A$: in the cross-section $x=1+O(\varepsilon)$, so that the source of Eq.~\eqref{eq:GSnorm}, $A+(1-A)x^2=1+O(\varepsilon)$, is constant at leading order. The reference equilibrium therefore solves
	\begin{equation}
		\frac{\partial^2u}{\partial x^2}+\frac{\partial^2u}{\partial y^2}=-1 \ \text{in}\ \Omega,\qquad
		u=0 \ \text{on}\ \partial\Omega,
		\label{eq:cyl}
	\end{equation}
	where $u$ is the poloidal flux of the cylinder, normalized so that it is positive inside the plasma. 
	%, where the poloidal field vanishes, 
	Its maximum is the magnetic axis of the cylindrical equilibrium and $x_\text{cyl}$ is its radial position. Equation~\eqref{eq:cyl} is the equilibrium of a straight current column with uniform current density whose boundary is a flux surface, as in the large-aspect-ratio model of non-circular tokamaks~\cite{strauss1976nonlinear}. 
It is solved as in Sec.~\ref{sec:method}, following~\cite{li2008}, with the particular solution $u_p=-[(x-1)^2+y^2]/4$ and the harmonic polynomials $\mathrm{Re}\,[(x-1+iy)^k]$~\cite{axler2001}, $k=0,\ldots,16$, in place of $\psi_p$ and of the basis of Eq.~\eqref{eq:decomp} (the imaginary parts are not needed, since the boundaries are up-down symmetric).
For a boundary symmetric in the radial direction, such as a circle or an ellipse, $x_\text{cyl}$ is the center and the classical definition is recovered. The Shafranov shift is then defined as
	\begin{equation}
		\frac{\Delta}{a}=\frac{x_\text{ax}-x_\text{cyl}}{\varepsilon},
		\label{eq:shift}
	\end{equation}
	a positive value indicating a magnetic axis on the low-field side of the reference, that is, at a larger major radius.
	Measuring the axis from any other point $x_\text{ref}$ gives
	\begin{equation}
		\frac{\Delta_\text{ref}}{a}=\frac{x_\text{ax}-x_\text{ref}}{\varepsilon}=\frac{\Delta}{a}+e_\text{ref},\qquad
		e_\text{ref}=\frac{x_\text{cyl}-x_\text{ref}}{\varepsilon},
		\label{eq:err}
	\end{equation}
	where $e_\text{ref}$ is the error introduced by the choice of the reference.
	Four alternatives are compared: the geometric center $x_\text{geom}$ of Eq.~\eqref{eq:shafranov}, with error $e_\text{geom}$, for which $\Delta_\text{ref}=\Delta_\text{geom}$ is the conventional shift; the barycenter of the LCFS, $x_\text{bary}=\oint x\,dl/\oint dl$; the area centroid, $x_A=\iint x\,dx\,dy/\iint dx\,dy$; and the current centroid, $x_J=\iint j_\varphi x\,dx\,dy/\iint j_\varphi\,dx\,dy$, with $j_\varphi$ from Eq.~\eqref{eq:jtor}. The first three depend only on the boundary, $x_J$ also on $A$.
	The reference $x_\text{cyl}$ is insensitive to the current profile assumed in the cylinder: replacing the uniform current by a peaked one changes it by less than $0.001\,\varepsilon$ for the boundaries tested.
	
	\section{Shaped tokamak boundaries: triangularity and squareness}
	\label{sec:convex}

	\textit{Family A: Miller boundaries.}	The Miller parametrization~\cite{miller1998},
	\begin{equation}
		x=x_0+\varepsilon\cos(\theta+\arcsin\delta\cdot\sin\theta), \quad
		y=\varepsilon\kappa\sin\theta,
		\label{eq:miller}
	\end{equation}
	where $\theta$ is a parametric angle (it coincides with the geometric poloidal angle only for $\delta=0$ and $\kappa=1$), $\kappa$ is the elongation and $\delta$ the triangularity, describes the progression from D-shapes ($\delta>0$) to NT plasmas
	($\delta<0$). Three cases are studied, with $x_0=1$, $\varepsilon=0.32$ and $\kappa=1.7$: a positive triangularity (PT) boundary (A1, $\delta=+0.5$), an ellipse (A2, $\delta=0$) and an NT boundary (A3, $\delta=-0.5$). The dependence on triangularity is then examined by a continuous scan in $\delta$ at fixed $\beta_p$ (Fig.~\ref{fig:dsweep}).
	
	\textit{Family B: Luce boundaries.}
	Squareness is introduced with the superellipse model of Luce~\cite{luce2013}, in which each quadrant of the boundary is a superellipse of exponent $n$ inscribed in the rectangle fixed by $\varepsilon$, $\kappa$ and $\delta$, so that the squareness $\zeta$ can be varied without changing the other shape parameters. The squareness measures, along the diagonal of each quadrant, how far the boundary lies outside ($\zeta>0$) or inside ($\zeta<0$) the ellipse with the same axes, and fixes the exponent $n$~\cite{luce2013}. The quadrants are elliptic for $\zeta=0$ ($n=2$), squarer for $\zeta>0$ and diamond-like for $\zeta<0$; they are straight for $\zeta=-1/\sqrt2\approx-0.707$ ($n=1$) and, extending the model to $n<1$, concave below. For $\delta=0$ and $\kappa=1$ the boundary is the four-fold star polygon of Eq.~\eqref{eq:sf} with the same exponent $n$ (Sec.~\ref{sec:nonconvex}), so that the squareness scan joins the shaped tokamak boundaries to the non-convex ones. For $\zeta<0$ ($n<2$) the vertices of the boundary on the midplane and at the top and bottom have infinite curvature, and for $\delta\neq0$ the curvature is discontinuous at the top and bottom points; these features are the source of the larger $\eta_\psi$ of Family~B and require harmonics of high order, so that Families~A and~B are computed with the basis through degree 12 ($N_b=13$). Five cases are studied, with $x_0=1$, $\varepsilon=0.32$ and $\kappa=1.7$: for $\delta=0$, a square boundary (B1, $\zeta=+0.4$), a diamond-like one (B2, $\zeta=-0.4$) and a concave one (B3, $\zeta=-0.8$); for $\zeta=-0.8$, the concave boundaries with $\delta=+0.5$ (B4) and $\delta=-0.5$ (B5). The dependence on squareness is then examined by a continuous scan in $\zeta$ at fixed $\beta_p$ for $\delta=0$ and $\pm0.5$ (Fig.~\ref{fig:zsweep}).
	
	\begin{figure}[htbp]
		\centering
		\includegraphics[width=\textwidth]{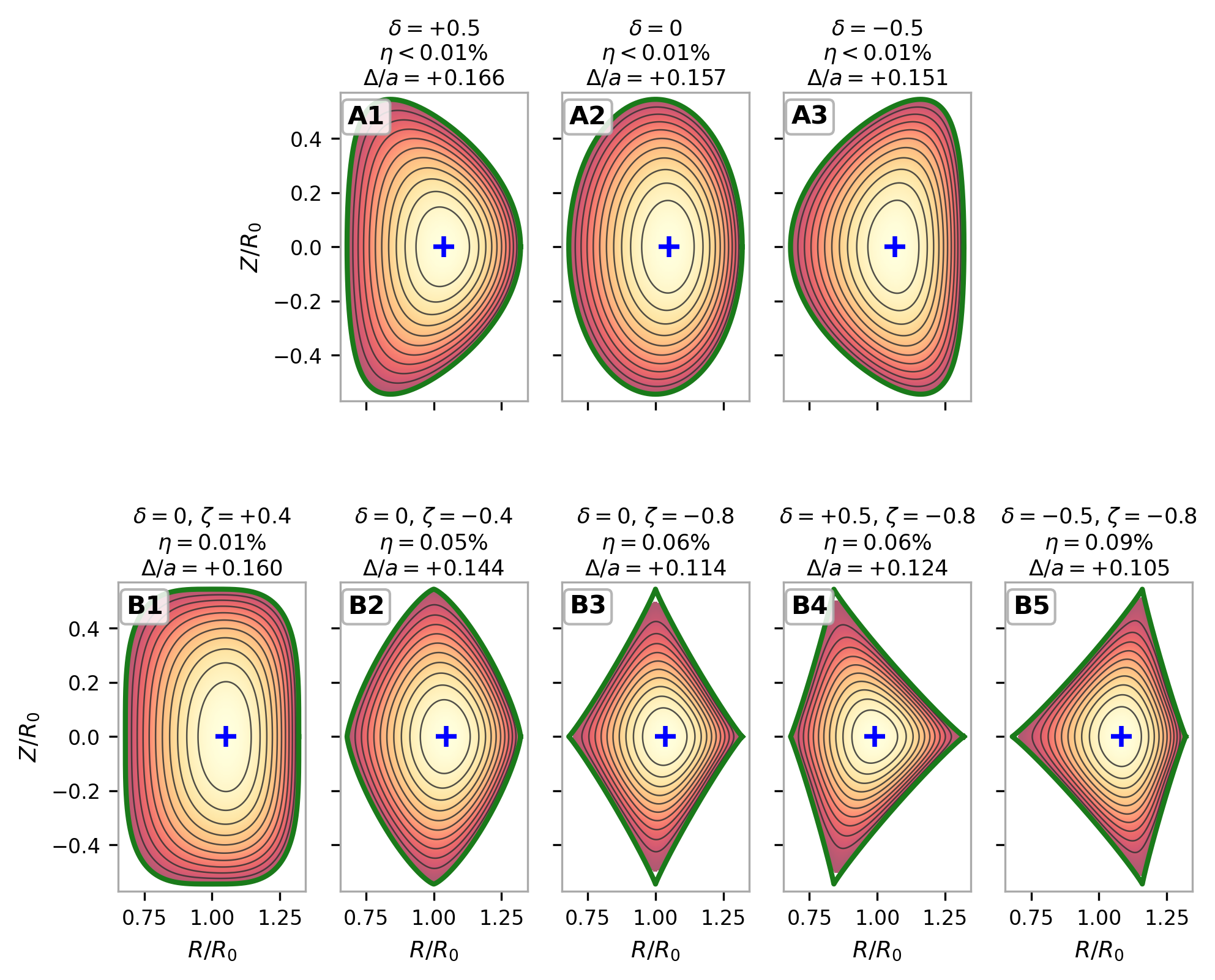}
		\caption{Family~A, top: Miller boundaries A1--A3, Eq.~\eqref{eq:miller}. Family~B, bottom: Luce boundaries B1--B5. All cases have $\kappa=1.7$, $\varepsilon=0.32$ and $A=0.11$; panel titles give $\delta$, $\zeta$, $\eta$ and $\Delta/a$, Eq.~\eqref{eq:shift}, and the figures of merit are listed in Table~\ref{tab:fom}. Background: $\tilde{p}=|\psihat|/|\psihat_\text{ax}|$, shown only inside the prescribed boundary; blue cross: magnetic axis; all panels share the same axes.}
		\label{fig:ABeq}
	\end{figure}
	
	\textit{Results.}
	All eight equilibria satisfy the criteria of Sec.~\ref{sec:method} ($\eta<0.1\,\%$, $\eta_\psi\leq6.5\,\%$) and have smooth nested flux surfaces (Fig.~\ref{fig:ABeq}); their figures of merit, computed at $A=0.11$ and $q_*=2$, are listed in Table~\ref{tab:fom}.

	The poloidal beta depends weakly on triangularity ($0.864\leq\beta_p\leq0.890$ for the Miller boundaries), since $j_\varphi$ depends only on the major radius, Eq.~\eqref{eq:jtor}. Negative squareness has a stronger effect: $\beta_p$ increases from $0.863$ for the square boundary B1 to $1.10$--$1.11$ for the concave boundaries B3--B5, as the ratio $C_p^2/\hat{V}$ in Eq.~\eqref{eq:betap} increases; the same mechanism raises $\beta_p$ of the star polygons (Sec.~\ref{sec:nonconvex}).
	
	The Shafranov shift, Eq.~\eqref{eq:shift}, is outward in all eight cases (Table~\ref{tab:fom}). For the Miller boundaries it decreases slightly from $\delta=+0.5$ to $\delta=-0.5$ ($\Delta/a=0.166$ and $0.151$), whereas the conventional shift increases from $\Delta_\text{geom}/a=0.114$ to $0.203$. The difference is the error of the geometric center, $e_\text{geom}$ (Table~\ref{tab:err}), defined in Eq.~\eqref{eq:err} with $x_\text{ref}=x_\text{geom}$. For positive triangularity the cross-section is wider on the inboard side, so that $x_\text{cyl}$ lies inboard of $x_\text{geom}$ ($e_\text{geom}<0$) and the conventional definition underestimates the shift; for negative triangularity the opposite occurs. For the Luce boundaries with $\delta=0$ the cross-section is symmetric in the radial direction, so that $x_\text{cyl}=x_\text{geom}$, and the shift decreases with the squareness, from $0.160$ (B1) to $0.114$ (B3). For the concave boundaries with $\delta=\pm0.5$ the error reaches $e_\text{geom}=\mp0.151$, and the conventional shift of B4 is negative, $\Delta_\text{geom}/a=-0.027$, although the axis is displaced outward ($\Delta/a=0.124$).
	
	Since $\beta_p$ varies with shape at fixed $A$, the role of triangularity is isolated by a scan of Miller boundaries at fixed $\beta_p=0.8$, with $A$ adjusted for each shape (Fig.~\ref{fig:dsweep}). For $\kappa=1.7$ the Shafranov shift (blue line) is nearly independent of triangularity, varying only from $\Delta/a=0.144$ ($\delta=-0.5$) to $0.156$ ($\delta=+0.5$), whereas the conventional shift (black line) decreases from $\Delta_\text{geom}/a=0.196$ to $0.104$. The same behaviour is found for $\kappa=1$, with $\Delta/a$ between $0.141$ and $0.176$ and $\Delta_\text{geom}/a$ decreasing from $0.263$ to $0.053$. Therefore, the dependence of the conventional shift on triangularity comes almost entirely from the error of the geometric center. Since $x_\text{geom}=1$ for all Miller boundaries, $e_\text{geom}=(x_\text{cyl}-1)/\varepsilon$ is set by the position of the cylindrical axis, which moves inboard as $\delta$ increases; the computed values are nearly linear in $\delta$, $e_\text{geom}\approx-0.25\,\delta$ for $\kappa=1$ and $-0.10\,\delta$ for $\kappa=1.7$ (for instance, $e_\text{geom}=-0.122$ and $-0.053$ at $\delta=+0.5$).
		
	For $\kappa=1$ and small $\delta$ the coefficient $-0.25$ follows from the Miller boundary, written about its geometric center as $r(\theta)/\varepsilon\approx1-(\delta/4)\cos\theta+(\delta/4)\cos3\theta$~\cite{song2021}, whose first harmonic, a rigid displacement by $-\delta\varepsilon/4$, moves $x_\text{cyl}$ by the same amount, so that $e_\text{geom}\approx-\delta/4$. In the equilibria of~\cite{song2021}, computed with a fixed LCFS, the larger shift of NT plasmas is attributed to this harmonic: triangularity changes the intercept of the linear dependence of the shift on $\beta_p+l_i/2$ but hardly its slope. This offset is the error of the geometric center, which depends only on the boundary and decreases in magnitude as $\kappa$ increases.

	These results are consistent with the CHEASE~\cite{chease1996} equilibria of TCV-like shapes with $|\delta|\approx0.5$ and identical profiles reviewed in~\cite{marinoni2021}, in which the shift is measured from the geometric center of the LCFS: it is below $5\,\%$ for PT and above $10\,\%$ for NT, and the NT shift remains larger than the high-$\beta$ PT one even at zero pressure (see Fig. 3 in \cite{marinoni2021}). The conventional shift of the present model shows the same behaviour (Fig.~\ref{fig:dsweep} and Table~\ref{tab:fom}). For $\kappa=1.7$ and $\beta_p=0.8$ it is $\Delta_\text{geom}/a=0.104$ for PT and $0.196$ for NT; at zero pressure the NT boundary (A3$_0$) still has $\Delta_\text{geom}/a=0.099$, comparable to the PT value at $\beta_p=0.8$.Instead, the PT boundary (A1$_0$) has $\Delta_\text{geom}/a=-0.004$.
	Measured from $x_\text{cyl}$, the same comparison gives a different result. For $\kappa=1.7$ and $\beta_p=0.8$ the shift is $\Delta/a=0.156$ for PT and $0.144$ for NT; at zero pressure the NT boundary (A3$_0$) has $\Delta/a=0.046$, less than a third of the PT value at $\beta_p=0.8$, and the PT boundary (A1$_0$) has $\Delta/a=0.048$. At the same pressure, the two signs of triangularity thus give nearly the same shift. The difference $\Delta_\text{geom}/a-\Delta/a$ is the error of the geometric center, $e_\text{geom}=-0.053$ for PT and $+0.053$ for NT (Table~\ref{tab:err}), which depends only on the boundary and is the same at any pressure: it lowers the conventional shift of PT and raises that of NT by the same amount. The larger shift of NT plasmas is therefore not an effect of triangularity on the displacement of the axis, but an error of the reference point. The axis of NT plasmas does lie closer to the outboard boundary, but this is already the case in a straight cylinder with the same cross-section.
	
	The other centers defined by the boundary do not perform better (Fig.~\ref{fig:dsweep}, Table~\ref{tab:err}). The barycenter and the area centroid follow $x_\text{cyl}$ for $\kappa=1$, but for $\kappa=1.7$ they overestimate the effect of triangularity: the shift measured from the barycenter varies by $0.212$ over the scan, more than twice the variation of $\Delta_\text{geom}/a$. The current centroid lies outboard of the area centroid in all cases, by $e_A-e_J=0.03$--$0.07$ (Table~\ref{tab:err}), because for $A=0.11$ the current density increases with $x$; it therefore lies outboard of $x_\text{cyl}$ ($e_J<0$) in all cases except A1, whose area centroid is far on the inboard side ($e_A=+0.071$, $e_J=+0.006$).
	
	\begin{figure}[htbp]
		\centering
		\includegraphics[width=1\textwidth]{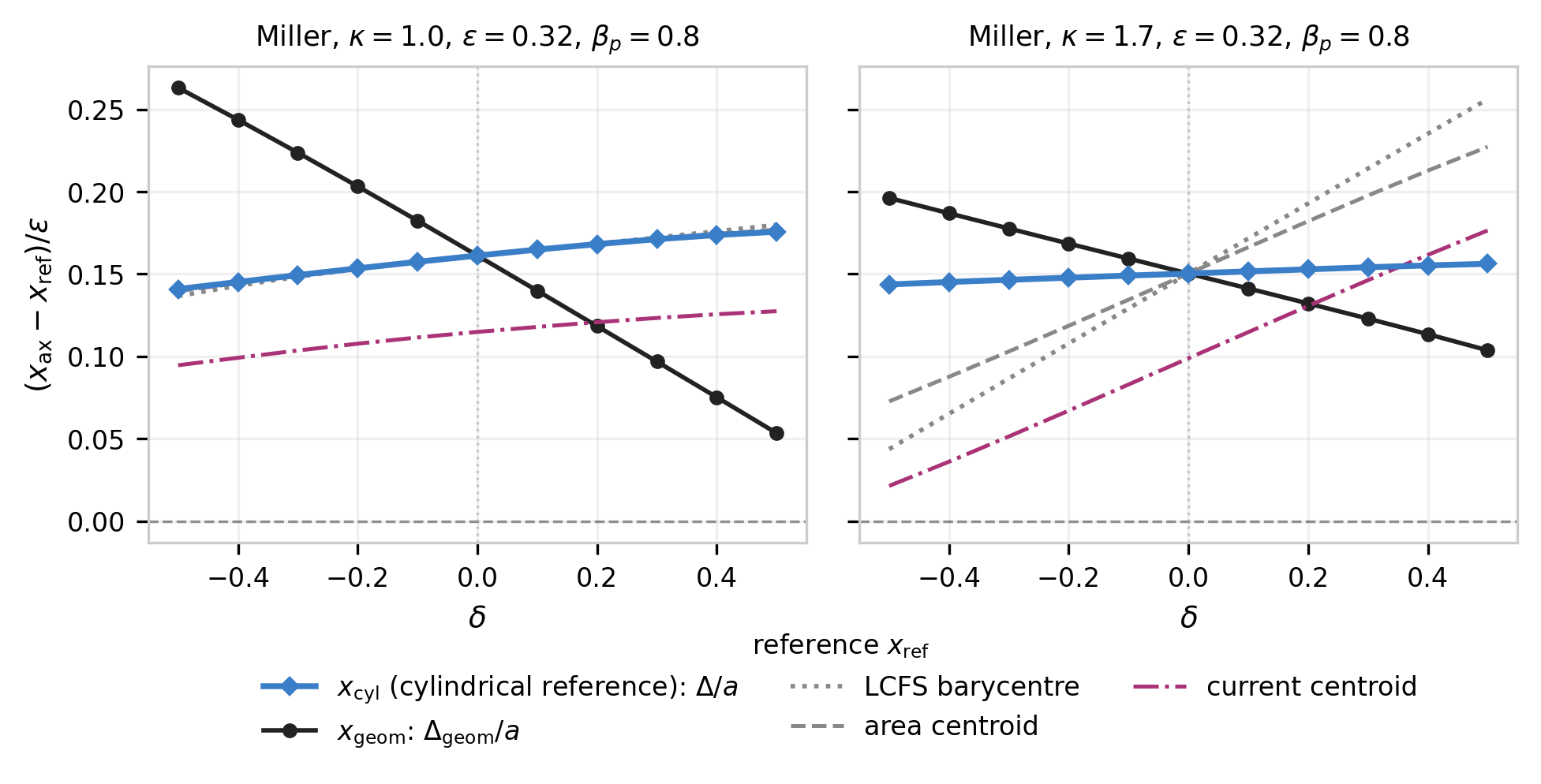}
		\caption{Miller boundaries ($\varepsilon=0.32$; $\kappa=1$ left, $\kappa=1.7$ right) at fixed $\beta_p=0.8$: axis position measured from the cylindrical reference $x_\text{cyl}$ ($\Delta/a$, blue), from $x_\text{geom}$ ($\Delta_\text{geom}/a$, black), from the LCFS barycenter (grey, dotted), the area centroid (grey, dashed) and the current centroid (purple, dash-dotted), versus triangularity.}
		\label{fig:dsweep}
	\end{figure}
	
	\begin{table}[tb!]
		\centering
\caption{Figures of merit at $q_*=2$ and $A=0.11$, except for A1$_0$ and A3$_0$, which have the boundaries of A1 and A3 at zero pressure ($A=1$).
	$\varepsilon=a/R_0$; $C_p=\oint dl/R_0$; $\hat{V}=\iint x\,dx\,dy$;
	$\beta_p$ is independent of $q_*$; $\beta_t=\beta_p\varepsilon^2/q_*^2$;
	$\Delta_\text{geom}/a$ from Eq.~\eqref{eq:shafranov}, Shafranov shift $\Delta/a$ from Eq.~\eqref{eq:shift}.
	$^\dagger$: $\eta_\psi>10\,\%$, Eq.~\eqref{eq:etapsi}; values approximate.}
		\label{tab:fom}
		\renewcommand{\arraystretch}{1.15}
		\resizebox{\textwidth}{!}{%
			\begin{tabular}{l c c c c c c c c}
				\toprule
				Geometry & $A$ & $\varepsilon$ & $C_p$ & $\hat{V}$ &
				$\beta_p$ & $\beta_t\,[\%]$ & $\Delta_\text{geom}/a$ & $\Delta/a$ \\
				\midrule
				\multicolumn{9}{l}{\textit{Family A --- Miller boundaries ($\kappa=1.7$, $\varepsilon=0.32$)}} \\
				A1 Miller $\delta=+0.5$ (positive triangularity)       & 0.11 & 0.320 & 2.782 & 0.507 & 0.890 & 2.28 & $+0.114$ & $+0.166$ \\
				A2 Miller $\delta=0$ (ellipse)                         & 0.11 & 0.320 & 2.760 & 0.547 & 0.864 & 2.21 & $+0.157$ & $+0.157$ \\
				A3 Miller $\delta=-0.5$ (NT)                           & 0.11 & 0.320 & 2.782 & 0.549 & 0.865 & 2.21 & $+0.203$ & $+0.151$ \\
				A1$_0$ Miller $\delta=+0.5$, zero pressure             & 1    & 0.320 & 2.782 & 0.507 & 0     & 0    & $-0.004$ & $+0.048$ \\
				A3$_0$ Miller $\delta=-0.5$, zero pressure             & 1    & 0.320 & 2.782 & 0.549 & 0     & 0    & $+0.099$ & $+0.046$ \\
				\midrule
				\multicolumn{9}{l}{\textit{Family B --- Luce boundaries ($\kappa=1.7$, $\varepsilon=0.32$)}} \\
				B1 Luce $\delta=0$, $\zeta=+0.4$ (square)              & 0.11 & 0.320 & 3.015 & 0.635 & 0.863 & 2.21 & $+0.160$ & $+0.160$ \\
				B2 Luce $\delta=0$, $\zeta=-0.4$ (diamond-like)        & 0.11 & 0.320 & 2.576 & 0.438 & 0.911 & 2.33 & $+0.144$ & $+0.144$ \\
				B3 Luce $\delta=0$, $\zeta=-0.8$ (concave)             & 0.11 & 0.320 & 2.529 & 0.321 & 1.095 & 2.80 & $+0.114$ & $+0.114$ \\
				B4 Luce $\delta=+0.5$, $\zeta=-0.8$ (concave)          & 0.11 & 0.320 & 2.589 & 0.302 & 1.105 & 2.83 & $-0.027$ & $+0.124$ \\
				B5 Luce $\delta=-0.5$, $\zeta=-0.8$ (concave)          & 0.11 & 0.320 & 2.589 & 0.339 & 1.107 & 2.83 & $+0.255$ & $+0.105$ \\
				\midrule
				\multicolumn{9}{l}{\textit{Family C --- Star-polygon scan ($\varepsilon=0.32$, $q_*=2$)}} \\
				Any $m$, $n=2$ (circle) & 0.11 & 0.320 & 2.011 & 0.322 & 0.900 & 2.30 & $+0.174$ & $+0.174$ \\
				C1 $m=2$, $n=1$ (lens) & 0.11 & 0.320 & 1.654 & 0.205 & 0.911 & 2.33 & $+0.153$ & $+0.153$ \\
				C2 $m=3$, $n=1\approx n^\star$ & 0.11 & 0.320 & 2.017 & 0.281 & 0.948 & 2.43 & $-0.014$ & $+0.158$ \\
				C3 $m=4$, $n=1$ (square) & 0.11 & 0.320 & 1.810 & 0.205 & 1.012 & 2.59 & $+0.107$ & $+0.107$ \\
				C4 $m=5$, $n=1$ & 0.11 & 0.320 & 2.117 & 0.250 & 1.096 & 2.81 & $+0.021$ & $+0.127$ \\
				C5 $m=6$, $n=1^\dagger$ & 0.11 & 0.320 & 2.032 & 0.205 & 1.201 & 3.07 & $+0.107$ & $+0.107$ \\
				C6 $m=7$, $n=1^\dagger$ & 0.11 & 0.320 & 2.272 & 0.227 & 1.331 & 3.41 & $+0.061$ & $+0.113$ \\
				$m=5$, $n=0.80\approx n^\star$$\,^\dagger$ & 0.11 & 0.320 & 2.249 & 0.193 & 1.418 & 3.63 & $-0.012$ & $+0.094$ \\
				$m=3$, $n=0.80$ & 0.11 & 0.320 & 2.103 & 0.249 & 1.071 & 2.74 & $-0.119$ & $+0.136$ \\
				$m=3$, $n=0.50$ (concave star)$^\dagger$ & 0.11 & 0.320 & 2.291 & 0.121 & 2.034 & 5.21 & $-0.275$ & $+0.059$ \\
				\bottomrule
		\end{tabular}}
	\end{table}
	
	\begin{table}[tb!]
		\centering
		\caption{Error $e_\text{ref}=(x_\text{cyl}-x_\text{ref})/\varepsilon$, Eq.~\eqref{eq:err}, of four alternative references, and Shafranov shift $\Delta/a$, Eq.~\eqref{eq:shift}, for the cases of Families~A and~B and representative cases of Family~C in Table~\ref{tab:fom}. The shift measured from $x_\text{ref}$ is $\Delta/a+e_\text{ref}$. $^\dagger$: $\eta_\psi>10\,\%$.}
		\label{tab:err}
		\renewcommand{\arraystretch}{1.15}
		\begin{tabular}{l c c c c c c}
			\toprule
			Case & $A$ & $e_\text{geom}$ & $e_\text{bary}$ & $e_A$ & $e_J$ & $\Delta/a$ \\
			\midrule
			A1 (Miller, $\delta=+0.5$)  & 0.11 & $-0.053$ & $+0.100$ & $+0.071$ & $+0.006$ & $+0.166$ \\
			A2 (Miller, $\delta=0$)     & 0.11 & $0$ & $0$ & $0$ & $-0.062$ & $+0.157$ \\
			A3 (Miller, $\delta=-0.5$)  & 0.11 & $+0.053$ & $-0.100$ & $-0.071$ & $-0.133$ & $+0.151$ \\
			B1 (Luce, $\zeta=+0.4$)     & 0.11 & $0$ & $0$ & $0$ & $-0.072$ & $+0.160$ \\
			B2 (Luce, $\zeta=-0.4$)     & 0.11 & $0$ & $0$ & $0$ & $-0.050$ & $+0.144$ \\
			B3 (Luce, $\zeta=-0.8$)     & 0.11 & $0$ & $0$ & $0$ & $-0.039$ & $+0.114$ \\
			B4 (Luce, $\delta=+0.5$, $\zeta=-0.8$) & 0.11 & $-0.151$ & $+0.048$ & $+0.029$ & $-0.014$ & $+0.124$ \\
			B5 (Luce, $\delta=-0.5$, $\zeta=-0.8$) & 0.11 & $+0.151$ & $-0.048$ & $-0.029$ & $-0.070$ & $+0.105$ \\
			\midrule
			C2 ($m=3$, $n=1$)         & 0.11 & $-0.172$ & $0$ & $0$ & $-0.057$ & $+0.158$ \\
			C3 ($m=4$, $n=1$)         & 0.11 & $0$ & $0$ & $0$ & $-0.041$ & $+0.107$ \\
			C4 ($m=5$, $n=1$)         & 0.11 & $-0.106$ & $0$ & $0$ & $-0.051$ & $+0.127$ \\
			$m=3$, $n=0.8$            & 0.11 & $-0.254$ & $0$ & $0$ & $-0.053$ & $+0.136$ \\
			$m=3$, $n=0.5^\dagger$    & 0.11 & $-0.333$ & $0$ & $0$ & $-0.032$ & $+0.059$ \\
			\bottomrule
		\end{tabular}
	\end{table}
	
		The role of squareness is isolated in the same way by a scan of Luce boundaries with $\kappa=1.7$, $\delta=0$ and $\pm0.5$, at fixed $\beta_p=0.8$ (Fig.~\ref{fig:zsweep}, top row; $\eta\leq0.13\,\%$, $\eta_\psi\leq7.8\,\%$). Unlike triangularity, squareness changes the Shafranov shift $\Delta$ of Eq.~\eqref{eq:shift} itself. For $\delta=0$ (Fig.~\ref{fig:zsweep}, bottom center, blue line) the shift is nearly independent of positive squareness but decreases for negative squareness, from $\Delta/a=0.150$ at $\zeta=0$ to $0.105$ at $\zeta=-0.7$ (nearly straight sides) and $0.081$ at $\zeta=-0.9$, in the shaded region where the quadrants are concave. The same decrease is found at fixed $A$ (B1--B3 in Table~\ref{tab:fom}), so that it is not a consequence of the change of the current profile. For $\delta=\pm0.5$ (Fig.~\ref{fig:zsweep}, bottom left and right) the shift (blue lines) is nearly the same as for $\delta=0$, whereas the conventional shift (black lines) moves away from it as the squareness decreases, because the error of $x_\text{geom}$ grows (up to $\mp0.171$ at $\zeta=-0.9$). For $\delta=+0.5$ the conventional shift even becomes negative below $\zeta\approx-0.64$ (bottom left), close to the onset of concavity, although the magnetic axis lies outboard of $x_\text{cyl}$ ($\Delta>0$), as for the three-fold star polygons of Sec.~\ref{sec:nonconvex}. Triangularity thus changes mainly the conventional shift, through the error of the geometric center, whereas squareness changes the Shafranov shift itself.
	
For $\delta=\pm0.5$, the distance from the blue curve varies much less along the scan for the barycenter and the area centroid (grey lines in Fig.~\ref{fig:zsweep}, bottom left and right) than for $x_\text{geom}$ (black), which nearly coincides with the blue curve at $\zeta=+0.5$ and moves away from it as the squareness decreases. The grey curves are therefore closer to the blue one for $\zeta\lesssim-0.2$, which includes the concave cases B4 and B5 ($|e_\text{bary}|=0.048$ and $|e_A|=0.029$, against $|e_\text{geom}|=0.151$, in Table~\ref{tab:err}), while above this value the black curve is the closer one. Together with the overestimation found for the Miller boundaries (Fig.~\ref{fig:dsweep}), this shows that no center defined by the boundary alone coincides with $x_\text{cyl}$ for all shapes.
	
%	Finally, Miller and Luce boundaries with the same triangularity are not the same shape: the Miller boundary with $\delta=+0.5$ and $\kappa=1.7$ has squareness $\zeta=-0.17$ on the outboard side and $+0.23$ on the inboard side, and the Luce boundaries with $\zeta=0$ and $\delta=\pm0.5$ have $e_\text{geom}=\mp0.028$ (Fig.~\ref{fig:zsweep} at $\zeta=0$), about half the Miller value (Fig.~\ref{fig:dsweep} at $\delta=\pm0.5$).
	\begin{figure}[htbp]
		\centering
		\includegraphics[width=1\textwidth]{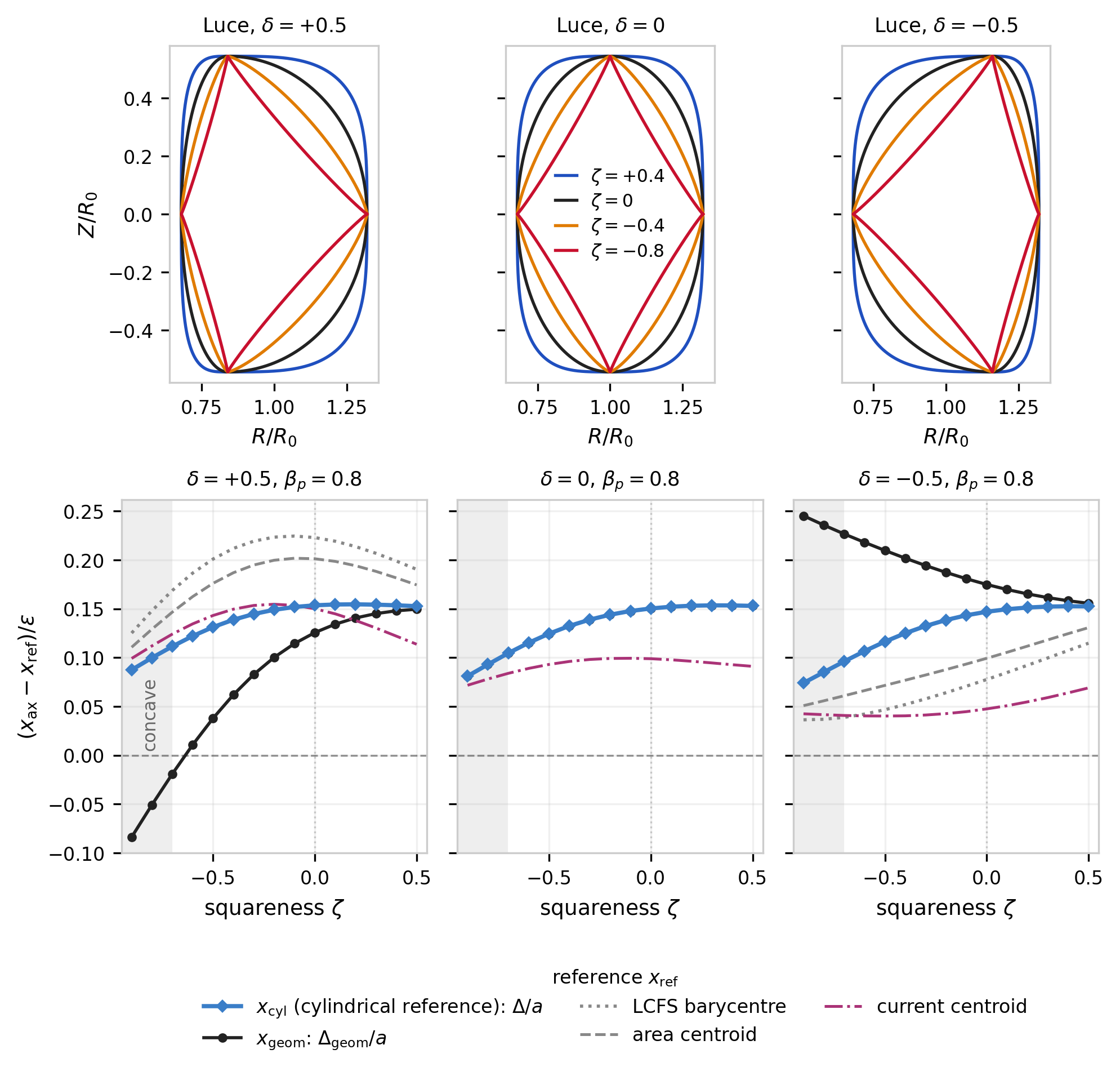}
		\caption{Luce boundaries ($\varepsilon=0.32$, $\kappa=1.7$) at fixed $\beta_p=0.8$. Top: boundaries with $\zeta=+0.4$, $0$, $-0.4$ and $-0.8$ for $\delta=+0.5$, $0$ and $-0.5$. Bottom: axis position measured from the cylindrical reference $x_\text{cyl}$ ($\Delta/a$, blue), from $x_\text{geom}$ ($\Delta_\text{geom}/a$, black), from the LCFS barycenter (grey, dotted), the area centroid (grey, dashed) and the current centroid (purple, dash-dotted), versus squareness. Shaded: concave quadrants ($\zeta<-1/\sqrt2$, $n<1$). For $\delta=0$ the boundary is symmetric in the radial direction and the first four references coincide. The vertical distance of each curve from the blue one is the error $e_\text{ref}$ of the corresponding reference, Eq.~\eqref{eq:err}.}
		\label{fig:zsweep}
	\end{figure}

	%% =============================================================================
	\section{Non-convex boundaries: the star-polygon scan}
	\label{sec:nonconvex}
	
\begin{figure}[p]
		\centering
		\includegraphics[width=0.86\textwidth]{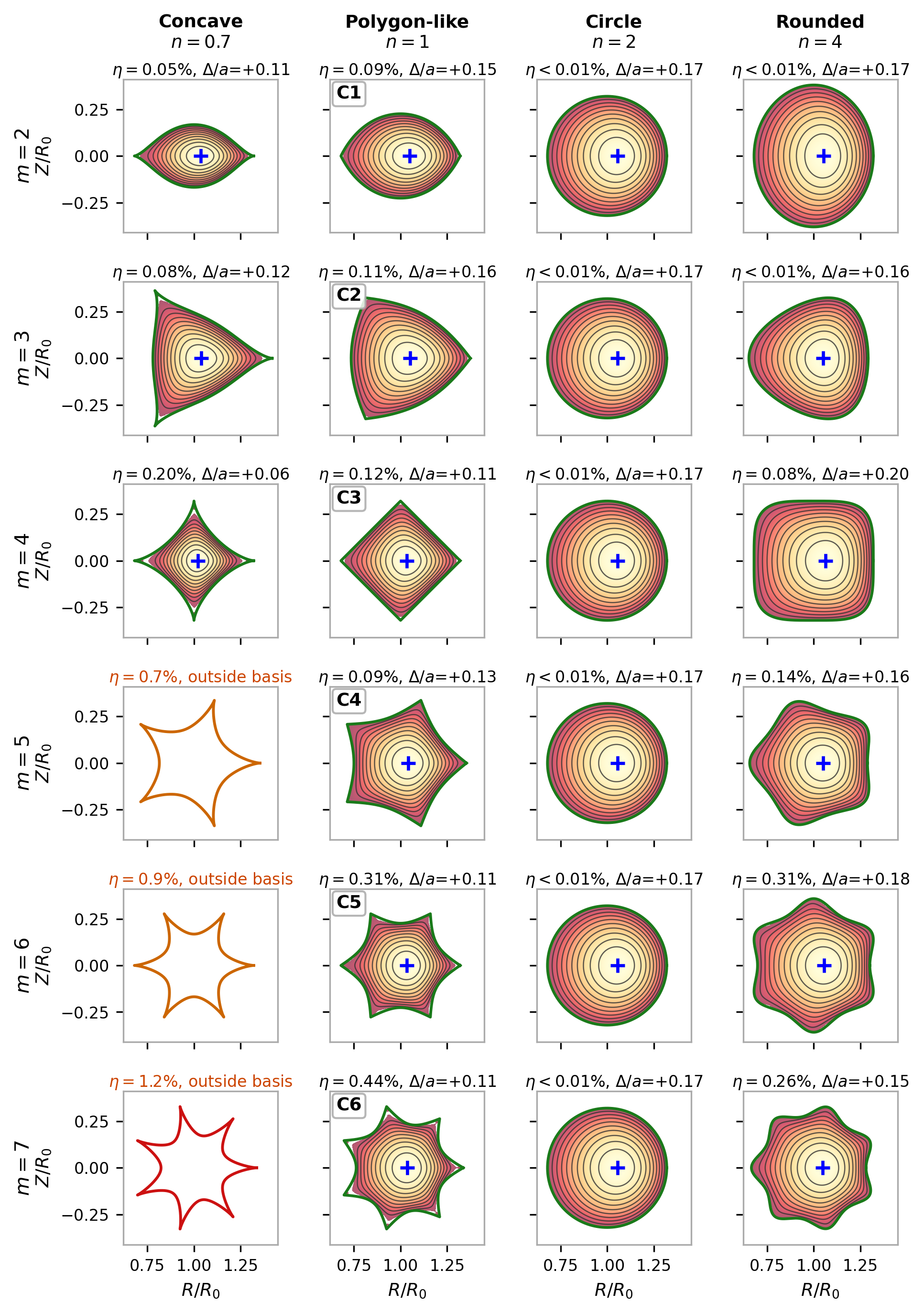}
		\caption{Family~C equilibria: $6\times4$ grid,
			rows $m\in\{2\ldots7\}$, columns $n\in\{0.7,\,1,\,2,\,4\}$.
			Green LCFS: $\eta<0.5\,\%$; orange/red: invalid. Panel titles give $\eta$ and $\Delta/a$, Eq.~\eqref{eq:shift}; all panels share the same axes.
			Labels C1--C6 on the polygon-like column ($n=1$).
			The $n=2$ column reduces to the same circle for all $m$
			($|\cos|^2+|\sin|^2\equiv1$).}
		\label{fig:Ceq}
	\end{figure}

	\textit{Family C: star-polygon boundaries.}
	The superformula of Gielis~\cite{gielis2003},
	\begin{equation}
		r(\theta)=\bigl[|\cos(m\theta/4)|^{n_2}+|\sin(m\theta/4)|^{n_3}
		\bigr]^{-1/n_1}s,
		\label{eq:sf}
	\end{equation}
	with Cartesian mapping
	\begin{equation}
		\begin{aligned}
			x &= x_0 + r(\theta) \cos\theta,\\
			y &= r(\theta) \sin\theta,
		\end{aligned}
		\label{eq:sf_cart}
	\end{equation}
	generates a boundary with $m$-fold rotational symmetry about $(x_0,0)$, here with $x_0=1$.
	With $n_1=n_2=n_3\equiv n$, as used here, $n<1$ gives a concave star with $m$ pointed vertices, $n=1$ a polygon-like boundary with $m$ corners (with straight sides only for $m=4$, a square with its vertices on the axes), and $n>1$ rounds the corners towards the circle, which is recovered at $n=2$ for every $m$.The scale factor $s$ is fixed so that $\varepsilon=(x_\text{max}-x_\text{min})/2=0.32$ exactly. The scan covers 54 cases, with $
	m\in\{2,3,4,5,6,7\}$ and $
	n\in\{0.50,\,0.65,\,0.80,\,1.0,\,1.2,\,1.5,\,2.0,\,3.0,\,5.0\}$, at $A=0.11$, with $N_b=9$ for $m=7$ and $N_b=7$ otherwise. Of these, 45 satisfy $\eta<0.5\,\%$ (green in Fig.~\ref{fig:Ceq}); the minimum exponent rises with $m$, from $n=0.50$ at $m=2$ to $n=1.0$ at $m=6$ and $7$, because sharper vertices and a higher symmetry order require harmonics beyond the angular resolution of the basis (Sec.~\ref{sec:solution}).
	
\begin{figure}[htbp]
		\centering
		\includegraphics[width=\textwidth]{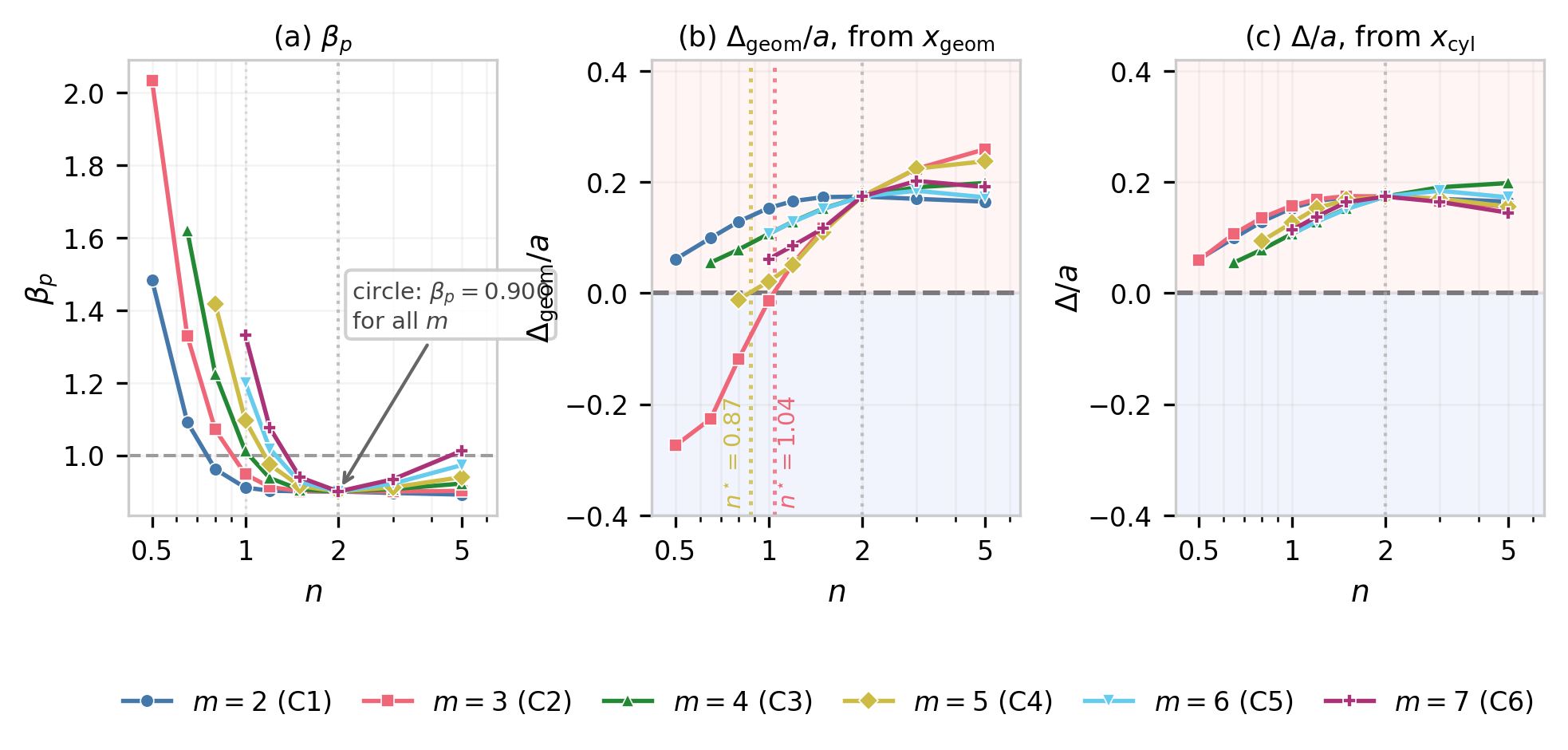}
		\caption{Family~C figures of merit vs $n$ (one curve per $m$,
			valid cases only).
			(a) $\beta_p(n)$; all curves cross at $n=2$
			($\beta_p=0.900$); the polygon-like cases C1--C6 of Fig.~\ref{fig:Ceq} and Table~\ref{tab:fom}, indicated in the legend, are the points at $n=1$.
			(b) Conventional shift $\Delta_\text{geom}/a(n)$, Eq.~\eqref{eq:shafranov}; dotted verticals at $n^\star(m=3)=1.04$
			and $n^\star(m=5)=0.87$, where $\Delta_\text{geom}/a$ changes sign for odd $m$.
			(c) Shafranov shift $\Delta/a(n)$, Eq.~\eqref{eq:shift}, positive for all valid cases.}
		\label{fig:Ccurves}
	\end{figure}

	\textit{Results.}
	At $n=2$ all six boundaries reduce to the same circle, with identical figures of merit ($\beta_p=0.900$, $\Delta/a=0.174$), which checks the consistency of the scan.
	The poloidal beta stays close to this value for $n\geq1.5$, between $0.89$ and $1.01$ (Fig.~\ref{fig:Ccurves}a), and increases as $n$ decreases below $1.5$ and the corners sharpen, the more so the larger $m$: at $n=1$ it increases from $0.911$ (C1) to $1.331$ (C6, Table~\ref{tab:fom}), and for the concave shapes it reaches $2.03$ at $(m=3,\,n=0.5)$. The rise follows the geometric ratio $C_p^2/\hat{V}$ in Eq.~\eqref{eq:betap}, which increases from $13.4$ (C1) to $22.8$ (C6) and to about $43$ at $(m=3,\,n=0.5)$: at fixed pressure and current a longer perimeter lowers the average poloidal field, $\bar{B}_p\propto1/C_p$~\cite{cerfon2010}.
	
\begin{figure}[htbp]
		\centering
		\includegraphics[width=\textwidth]{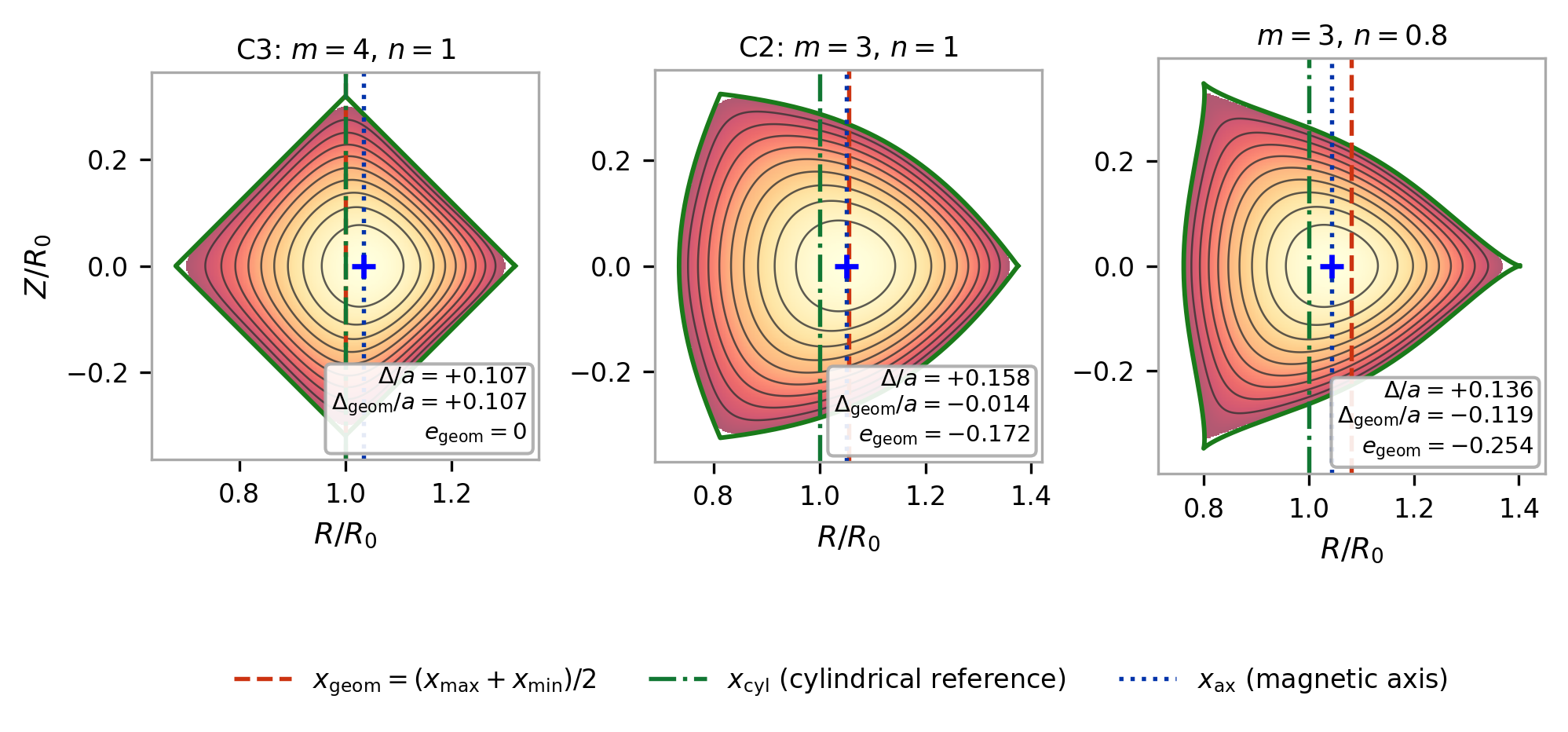}
		\caption{Reference points for C3 ($m=4$, $n=1$), C2 ($m=3$, $n=1$) and $(m=3,\,n=0.8)$: geometric center $x_\text{geom}$ (red, dashed), cylindrical reference $x_\text{cyl}$ (green, dash-dotted) and magnetic axis $x_\text{ax}$ (blue, dotted, and cross), with $\Delta/a$, $\Delta_\text{geom}/a$ and $e_\text{geom}$. For $m=4$, $x_\text{cyl}=x_\text{geom}$; for $m=3$, $x_\text{geom}$ lies outboard of the axis although the axis is displaced outward from $x_\text{cyl}$.}
		\label{fig:centers}
	\end{figure}

	The conventional shift (Fig.~\ref{fig:Ccurves}b) is positive for even $m$, as for the radially symmetric boundaries of Sec.~\ref{sec:convex}, but for $m=3$ and $5$ it changes sign at $n^\star(m=3)=1.04$ and $n^\star(m=5)=0.87$; for instance, $\Delta_\text{geom}/a=-0.014$ for C2 ($m=3$, $n=1$).
	The magnetic axis, however, lies on the low-field side of $R_0$ in every valid case, and the sign change is an error of the reference.
	Because of the $m$-fold rotational symmetry about $(1,0)$, $x_\text{cyl}=1$ for every $m$ and $n$, as do the LCFS barycenter and the area centroid (Table~\ref{tab:err}).
	The midpoint $x_\text{geom}$, instead, equals 1 only for even $m$. For odd $m$ and $n\leq1$ the outermost point is the vertex at $\theta=0$, where Eqs.~\eqref{eq:sf} and~\eqref{eq:sf_cart} give $r=s$ and $x_\text{max}=1+s$. The innermost point is either the midpoint of the opposite side, at $\theta=\pi$, where $m\theta/4$ is an odd multiple of $\pi/4$, so that $|\cos(m\theta/4)|=|\sin(m\theta/4)|=2^{-1/2}$ and Eq.~\eqref{eq:sf} gives $r=2^{1/2-1/n}s$, or the two vertices nearest to it, at $\theta=\pi\pm\pi/m$, where $r=s$ and $x=1-s\cos(\pi/m)$. Hence $x_\text{min}=1-\rho s$ and
	\begin{equation}
		e_\text{geom}=\frac{1-x_\text{geom}}{\varepsilon}=-\frac{1-\rho}{1+\rho},\qquad
		\rho=\max\left[\cos\frac{\pi}{m},\,2^{1/2-1/n}\right],
		\label{eq:geo_closed}
	\end{equation}
	with $e_\text{geom}=0$ for even $m$; for example, $e_\text{geom}=-0.172$ for C2 and $-1/3$ for $m=3$ and $n\leq2/3$. %The numerical values agree with Eq.~\eqref{eq:geo_closed} to within $10^{-6}$.
	 Figure~\ref{fig:centers} shows $x_\text{geom}$, $x_\text{cyl}$ and $x_\text{ax}$ for three cases: for $m=3$, $x_\text{geom}$ lies outboard of the magnetic axis, although the axis is displaced outward from $x_\text{cyl}$.
	
	The Shafranov shift, $\Delta/a=(x_\text{ax}-1)/\varepsilon$ (Fig.~\ref{fig:Ccurves}c), is positive in all 45 valid cases, from $0.055$ to $0.198$ (from $0.030$ to $0.182$ at fixed $\beta_p=0.8$).
	Three conclusions follow. First, the negative values of $\Delta_\text{geom}$ for odd $m$ are the error of $x_\text{geom}$, known in closed form from the boundary: the conventional shift becomes negative where $|e_\text{geom}|>\Delta/a$. Second, measured from the reference state the shift is outward for every shape and for both parities of $m$; there is no parity effect in the displacement of the axis. Third, the shift depends on shape through concavity: it decreases from $0.158$ (C2) to $0.059$ at $(m=3,\,n=0.5)$, and from $0.107$ (C3) to $0.055$ at $(m=4,\,n=0.65)$, as it decreases with negative squareness (Sec.~\ref{sec:convex}), of which the four-fold polygons are the case $\kappa=1$, $\delta=0$.
	
		\section{Conclusions}
	\label{sec:conclusions}
	
	The Shafranov shift has been studied with exact Solov'ev equilibria for a wide range of prescribed boundaries: Miller boundaries from positive to negative triangularity, Luce boundaries from positive to negative squareness, and star polygons of arbitrary symmetry order and concavity. The results concern the dependence of the poloidal beta on shape, the reference from which the shift is measured, and the dependence of the shift on shape.
	
	At fixed profiles, $\beta_p$ is nearly insensitive to triangularity ($0.864\leq\beta_p\leq0.890$) but increases with negative squareness, with the number of corners and with concavity, up to $2.03$ for the concave three-fold star, following the growth of the squared perimeter relative to the volume, $C_p^2/\hat{V}$. For this reason the shift has also been compared at fixed $\beta_p$.
	
	For shaped boundaries, the Shafranov shift depends on the point from which it is measured. The magnetic axis of the straight-cylinder equilibrium with the same cross-section, $x_\text{cyl}$, provides a zero-shift reference state: it depends only on the boundary and is obtained from the equilibrium of a straight conductor with uniform current density, Eq.~\eqref{eq:cyl}, without solving the toroidal equilibrium. Any other reference adds an error that also depends only on the boundary. For the geometric center this error is nearly linear in triangularity ($-\delta/4$ for $\kappa=1$) and is known in closed form for the star polygons, Eq.~\eqref{eq:geo_closed}. It makes the conventional shift negative for three- and five-fold star polygons with sharp vertices and for Luce boundaries with positive triangularity and nearly straight or concave quadrants, although the magnetic axis is always displaced outward. The barycenter and the area centroid of the boundary coincide with $x_\text{cyl}$ for rotationally symmetric shapes, but not in general: they overcorrect the effect of triangularity on elongated D-shapes and improve on $x_\text{geom}$ only for markedly negative squareness. No center defined by the boundary alone can therefore replace $x_\text{cyl}$.
	
	Measured from $x_\text{cyl}$, the shift is outward for every shape considered. At fixed $\beta_p$ it depends only weakly on triangularity ($\Delta/a=0.156$ and $0.144$ for $\delta=+0.5$ and $-0.5$ at $\kappa=1.7$), whereas the conventional shift is larger for negative triangularity, as in numerical equilibria~\cite{marinoni2021,song2021}. The larger shift of NT plasmas is therefore not an effect of triangularity on the displacement of the axis, but an error of the geometric center, present also at zero pressure. Squareness and concavity, instead, change the shift itself: for $\kappa=1.7$ it decreases from $\Delta/a=0.150$ for the ellipse to $0.081$ for concave quadrants, nearly independently of triangularity and both at fixed $\beta_p$ and at fixed profiles. The shift of the star polygons likewise decreases with concavity for every symmetry order, with no parity effect.
	
	The equilibria, with the shift measured from $x_\text{cyl}$, can also serve as benchmarks of shape effects for fixed-boundary~\cite{chease1996,helena1991} and free-boundary~\cite{lao1985,moret2015,fge2026} equilibrium codes, whose output is used for MHD stability, heating, current-drive and transport studies.
	
	\bibliography{refs}
	\bibliographystyle{unsrt}
	
\end{document}